\documentclass[aps,prl,twocolumn,showpacs,amsmath,floatfix,citeautoscript]{revtex4}
\usepackage{graphicx}
\usepackage{dcolumn}

\begin{document}

\title{Interface induced giant magnetoelectric coupling in 
multiferroic superlattices}

\author{Hongwei Wang}
\affiliation{Key Laboratory of Quantum Information,
University of Science and Technology of China, Hefei, Anhui, 230026, People's
Republic of China}

\author{Lixin He}
\email{helx@ustc.edu.cn}
\affiliation{Key Laboratory of Quantum Information,
University of Science and Technology of China, Hefei, Anhui, 230026,  People's
Republic of China}

\author{Xifan Wu }
\email{xifanwu@temple.edu}
\affiliation{Department of Physics and Institute for Computational Molecular Science
Temple University, Philadelphia, PA 19122, USA}
\date{\today}
\begin{abstract}

The electric and magnetic properties of (BaTiO$_3$)$_n$/(CaMnO$_3$)$_n$ 
short-period superlattices are studied by the first-principles calculations. 
The local electric polarizations in the CaMnO$_3$ layers are significant, 
comparable to that in the BaTiO$_3$ layers. 
Remarkably, the electric polarization 
is almost doubled when the spin configuration changes 
from antiferromagnetic to ferromagnetic in the superlattices,
indicating a giant magnetoelectric coupling. This enhancement of the magnetoelectric coupling is due to
the suppression of the antiferrodistortive mode in the CaMnO$_3$ layers at the interfaces.

\end{abstract}


\maketitle

The emergence of new phenomena at artificial heterostructure
is currently at the center of scientific and technological interest 
\cite{bousquet08,lee05,ohtomo04,yu10,rondinelli08}.
Because a large variety of degrees of freedom such as spin, charge, structural orderings
can be found in ABO$_3$ perovskite, it presents an ideal playground to explore the 
interactions among various orderings that could potentially lead to enhanced
functionalities, and multifunctional materials~\cite{bousquet08,rondinelli08}.

Multiferroics are among these multifunctional materials that have
attracted intensive interests recently~\cite{spaldin10}. 
The ferroelectric/(anti)-ferromagnetic superlattices (SLs) may have strong 
electric polarization, magnetic ordering and 
magnetoelectric coupling simultaneously which is rarely seen in single phase
bulk materials because of the symmetry restrictions~\cite{fiebig05}.
Conventionally, the magnetoelectric coupling 
can be introduced by the mechanical boundary condition 
through the in-plane strain or the electric boundary condition through 
charge continuity satisfied by the constituent materials~\cite{fiebig05}.
However, the interfaces separating different constituent materials can carry 
distinct instabilities originated from individual parent bulk materials~\cite{bousquet08}.
We show that the interfacial competition of instabilities \cite{bilc06} 
may lead to surprisingly enhancement of the magnetoelectric coupling, which arising from the
interfacial atomistic effects, can 
be much larger than that through the continuum media coupling.

AMnO$_3$ (A=Ca, Sr, ...) are good candidates as building blocks for the multiferroic
supperlattices, because they have several competing instabilities
\cite{bhattacharjee09} 
coupled to the magnetic ordering.
In AMnO$_3$, Mn atom has partially occupied $d$-orbitals, and 
the Mn-O-Mn bond angle is crucial in controlling 
the magnetic interactions, which at the same time also strongly couples to the 
oxygen octahedral rotation (AFD) and ferroelectric (FE) soft mode.
The coupling between the spin, the AFD modes and the FE modes depends on the relative 
energetics of these instabilities. Unfortunately in bulk AMnO$_3$,
there is a strong AFD instability associated with a large oxygen octahedral
rotation that suppresses the FE mode\cite{bhattacharjee09}.
As a result, the {\it linear} magnetoelectric (spin-ferroelectricity) 
coupling is usually found to be weak. 

In this letter, we demonstrate, through first-principles calculations, 
that the energetics of the AFD, FE and magnetic ordering can be drastically 
modified \cite{wu11} by interface engineering to favorite the magnetoelectric coupling.
We take the CaMnO$_3$(CMO)/BaTiO$_3$(BTO) SLs as our model systems.
We find that the MnO$_6$ octahedral rotation 
will be strongly suppressed by the neighboring BaO layers, leading to 
the enormous enhancement of magnetoelectric coupling as well as FE at interfaces.
This enhancement will be strengthened with the increasing density of
interfaces and reaches its maximum at the shortest SL, (i.e., $n$=1) 
where one observes a huge change of electric polarization between the 
antiferromagnetic (AFM) and ferromagnetic
(FM) states. To better clarify the mechanism for its underlying physics, an effective 
Hamiltonian model is also developed. The model explicitly shows 
that an increased magnetoelectric coupling
develops with a completely suppressed AFD, 
which consistently explains our computational results.

The calculations are based on the standard density functional
(DFT) theory with spin-polarized generalized gradient approximation,
implemented in the Vienna ab initio simulations package (VASP) 
\cite{kresse93,kresse96}. We adopt the
Perdew-Burke-Ernzerhof functional revised for solids (PBEsol)\cite{perdew08}. The
on-site Coulomb $U$=4.0 eV and exchange interaction $J$ =0.88 eV are
used for the Mn 3d electrons \cite{liechtenstein95}.
We use the plane-wave basis and projector augmented-wave
pseudopotentials \cite{blochl94}. A 500 eV energy
cutoff and 6$\times$6$\times$2 k-point meshes converge the results
very well.
All ionic coordinates are fully relaxed until
the Hellman-Feynman forces on the ions are less than 1 meV/{\AA}.
The electric polarizations are computed using the Berry phase
theory of polarization \cite{king-smith93}.


The calculated lattice constants of cubic CaMnO$_{3}$
and BaTiO$_{3}$ are 3.731{\AA} and 3.987{\AA}
respectively using PBEsol functional \cite{perdew08}, 
compared with the experimental
values 3.73{\AA} and 3.993{\AA}, which are much more
accurate than the those obtained from LDA. This is very important for
studying ferroelectrics, whose properties are very sensitive to the lattice
constants.

The ground state of orthorhombic (bulk) CaMnO$_{3}$ is G-type
antiferromagnetic. We find a small FE instability related to an
unstable polar mode  in the high symmetry cubic phase,
with an imaginary frequency $\omega_{FE}$=3.43$i$ cm$^{-1}$ 
and a much larger antiferro-distortive (AFD) instability
associated to a nonpolar oxygen rotational mode with an imaginary
frequency $\omega_{AFD}$=219$i$ cm$^{-1}$, consistent with previous
calculations\cite{bhattacharjee09}. As a result, the octahedra rotate a angle
about 8.94$^{\circ}$. In contrast, BaTiO3 is highly resistant to
oxygen octahedral rotations and exhibits a robust FE state at room
temperature.

The lattice mismatch between CMO and BTO is about 6.59\%, which
might be too large to grow high quality CMO/BTO SLs. One
may grow the CMO/BTO SLs on the NdGaO$_{3}$ substrate to
reduce the lattice mismatch between CMO and BTO to about
3.37\%.
We therefore fix the in-plane lattice constants
of the SL to those of NdGaO$_{3}$ substrates (3.86 \AA) in
the calculations. We have studied two short-period  SLs
structures (BaTiO$_3$)$_1$/(CaMnO$_3$)$_1$ (1:1) and
(BaTiO$_3$)$_2$/(CaMnO$_3$)$_2$ (2:2). In the calculations, we
fix the symmetry of the SLs to the space group P4bm,
allowing the MnO$_{6}$ and TiO$_{6}$ octahedra to rotate. We neglect
the tilting of MnO$_{6}$ octahedra (rotations about an in-plane
axis), because oxygen tilting requires a coherent pattern of tilts
that would propagate into the BTO unit cells, where octahedral
rotations are unfavorable.

\begin{figure}
\includegraphics[width=2.8in]{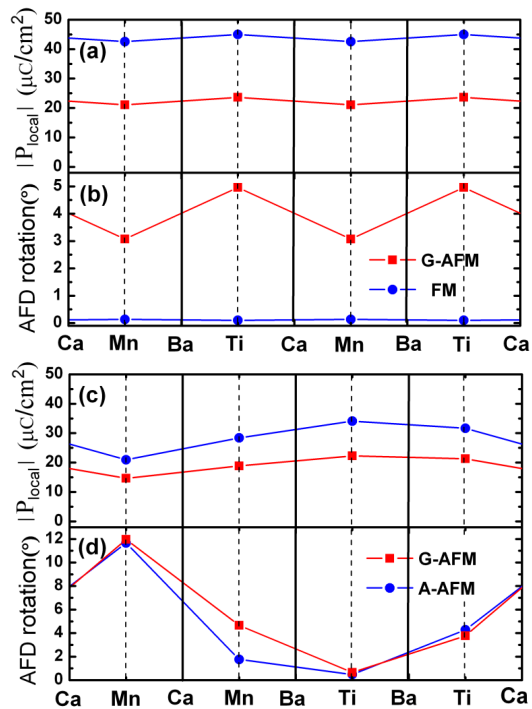}
\caption {\label{fig:fig1}(Color oline) (a) The local electric polarizations and (b) the AFD rotations
of the MnO$_2$ and TiO$_2$ layers in the 1:1 SL of the G-type AFM (red line) and FM
states (blue line).  (c) The local electric polarizations and (d) the AFD rotations
of the MnO$_2$ and TiO$_2$ layers in the 2:2 SL of the G-type AFM (red line) and
A-type AFM states (blue line). 
}
\end{figure}

To determine the ground-state spin structure, we calculate the total
energies of a set of spin configurations (SCs) 
for the 1:1 and 2:2 SLs with full
relaxations of the (electronic and lattice) structures. 
The results show that the stablest SCs are G-type AFM for 1:1, 
and A-type AFM for 2:2 SLs.
The calculated magnetic moments of Mn ions are about 2.9 $\mu_B$, 
whereas Ti ions have negligible induced magnetic moments.  We further fit the
exchange integrals to a Heisenberg model
$H=1/2\sum_{i,j} J_{ij} {\bf S}_i \cdot {\bf S}_j$, assuming nearest
neighbor coupling between the Mn ions.
In the 1:1 SL, the intra-layer exchange interaction 
$J_{\rm intra}$ = -4.8 meV/$\mu^{2}_{B}$, and the inter-layer 
exchange interaction 
$J_{\rm inter}$ = -0.86 meV/$\mu^{2}_{B}$.
In the 2:2 SL, the exchange interaction in the MnO$_2$ layer that sandwiched by the
two CaO layer $J^{(1)}_{\rm intra}$ = 8.61
meV/$\mu^{2}_{B}$, 
and in the MnO$_2$ layer that adjacent to the BaO layer $J^{(2)}_{\rm intra}$ = -1.11
meV/$\mu^{2}_{B}$. The interlayer exchange interaction $J^{(1)}_{\rm inter}$ =
-24.54  meV/$\mu^{2}_{B}$ and $J^{(2)}_{\rm inter}$ =-0.04  meV/$\mu^{2}_{B}$
respectively. Here, 
the magnetic moments ${\bf S}$ are normalized to 1. 
The $J^{(2)}_{\rm intra}$ in the 2:2 SL is slightly frustrated. The N\'{e}el
temperatures of the 1:1 and 2:2 SLs are 52 K and 59 K respectively calculated
from Monte Carlo simulations \cite{cao09}. This is
because the  N\'{e}el temperature of bulk CMO is low ($\sim$ 130 K)
\cite{wollan55}. The N\'{e}el temperature of the SLs can be enhanced by using
other AMnO$_3$ compounds with higher  N\'{e}el temperatures.

Figure \ref{fig:fig1} (b), (d), depict the local AFD associated with
the TiO$_6$ or MnO$_6$ octahedral rotation in 1:1 and 2:2 SLs respectively.
Clearly, the MnO$_6$ octahedral that sandwiched between two
CaO layers in 2:2 SL keeps a large rotation angle around 12$^\circ$ and 
TiO$_6$ octahedra that sandwiched between two BaO layers has a rotation almost 
around 0$^\circ$. This is in consistent with the fact that CMO bulk has a strong
AFD instability and BTO strongly resists the octahedra rotation. What is more
intriguing, however, is the behavior of the interfacial layers of MnO$_6$ octahedra.
Because of the presence of BaO on one side and CaO on the other side, MnO$_6$
at interface is exposed to a strongly broken mirror symmetry. As a result, 
this MnO$_6$ octahedra is found to have significantly reduced rotation angle which is 
around 4.6$^\circ$ and 3.0$^\circ$ in 2:2 and 1:1 SL respectively.

\begin{figure}
\centering
\includegraphics[width=2.8in]{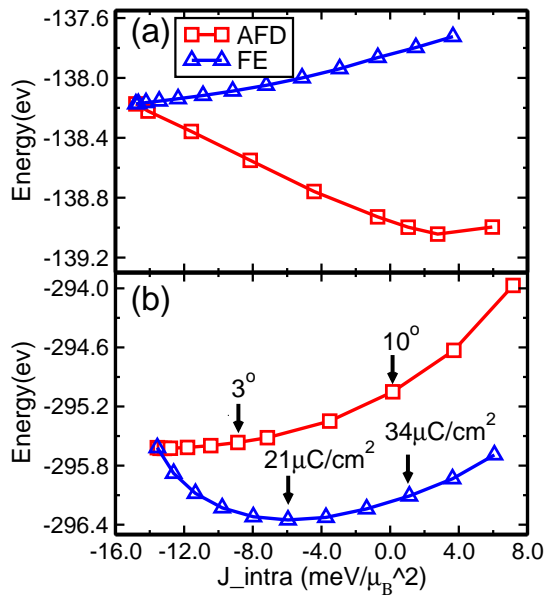}
\caption {\label{fig:exchange}(Color oline)
The total energies as functions of the MnO$_2$ intra-layer exchange energy
$J_{\rm intra}$ for (a) bulk CaMnO$_3$ and (b) 1:1 SL through AFD rotation
and FE modes. We also show the AFD rotation angle, and the
polarization of the FE mode in the 1:1 SL. } 
\end{figure}

In bulk CaMnO$_3$, it is the strong AFD that suppresses the FE instability from condensation at 
ground states. Because of the greatly reduced interfacial MnO$_6$ octahedra
rotation, together with strong polarization in the BaTiO$_3$ layers,
one might expect that the FE can develop in the CaMnO$_3$ layer of the SL.
Indeed, we find that both 1:1 and 2:2 SLs become multiferroic with  
spontaneous polarizations of 21.30 $\mu$C/cm$^2$ and 27.52 $\mu$C/cm$^2$ respectively computed
by Berry phase formalism.
To gain more insight, we calculate the layer polarization along [001] direction
$p_z=\sum_i{(e/\Omega) Z^{*}_i\lambda_i}$ based on the linear approximation involving
effective charges $Z^*_i$ and small ionic distortions $\lambda_i$ of each atoms $i$ 
in the cell from a higher central symmetric nonpolar reference structure.
The effective charges are obtained from the cubic CaMnO$_3$ and BaTiO$_3$ phase by 
finite difference method. The resulting layer polarizations are presented in Fig.1.
It can be seen that MnO$_2$ centered layers becomes polarized 
to almost the same degree of TiO$_2$-centered layers in both 1:1 and 2:2 SLs.

The development of multiferroic behavior in the CaMnO$_3$ component in the sperlattices has 
now been established. It provides us an opportunity to study the magnetoelectric coupling 
directly. To this end, we compare the electric polarizations under various spin configurations.
As one of the most interesting results, we find that the electric polarization undergoes 
a huge increase from about 21 $\mu$C/cm$^2$ to about 38 $\mu$C/cm$^2$ in 1:1 SL when the MnO$_2$ intra-layer
spin configuration is changed from antiferromatic  ($C$-type or $G$-type AFM) to ferromagnetic
(FM or $A$-AFM). In 2:2 SL, we observe a similar however weaker enhancement 
of polarization from about 19 $\mu$C/cm$^2$ to about 30 $\mu$C/cm$^2$.
The above result indicates that the magnetoelectric coupling is giant!

At a closer inspection of what happen locally in the SLs, the above drastic polarization change 
is also found to be accompanied by a almost homogeneous increased layer polarization for both MnO$_2$-centered 
and TiO$_2$-centered layers in both 1:1 and 2:2 SLs and a further drop of the interfacial 
MnO$_6$ octhedra rotations. 

It is well known that the Mn-O-Mn exchange angle is essential 
in controlling the magnetic ordering \cite{yamauchi08}.
In multiferroics where spin, FE and AFD instabilities are all present, 
Mn-O-Mn angle can be strongly coupled to MnO$_6$ octahedral rotation and FE soft phonon modes.
In Fig.~\ref{fig:exchange}(a), (b), we present the change of the total energy as 
a function of the intra-layer exchange integral $J_{\rm intra}$ in 
bulk CMO and 1:1 SL respectively. It can be seen that intra-layer spin ordering
can be continuously tuned from AFM($J_{\rm intra}<0$) to FM($J_{\rm intra}>0$) by
increasing either the MnO$_6$ octahedral rotation (red) or the amplitude of the 
FE soft phonon mode in both bulk CaMnO$_3$ and 1:1 SL.  In bulk CMO, the intra-layer AFM-FM phase transition 
is realized by the increased octahedral rotation which has a much lower total energy
than that of increased FE soft phonon mode. However in 1:1 SLs, the MnO$_6$ octahedral rotation
is lagerly suppressed by the interface because it
becomes energetically unfavorable [see Fig.~\ref{fig:exchange}(b)]. Thus, AFM-FM phase transition can be only driven 
by the increased FE soft phonon mode which results in the observed enhanced 
magnetoferroelectric coupling.

\begin{figure}
\centering
\includegraphics[width=2.8in]{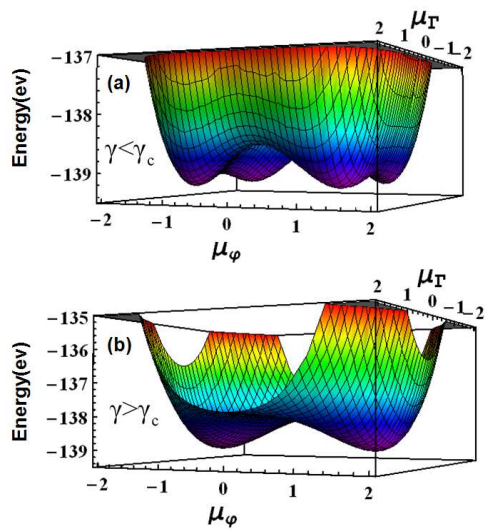}
\caption {Schematic shows of the energy surfaces of the effective Hamiltonian
  Eq.~\ref{eq:Heff}, in
  the case of (a) the coupling coefficient $\gamma < \gamma_c$, and 
(b) the coupling coefficient $\gamma > \gamma_c$, when both $\alpha <$0, and
  $\beta <$0.}
\label{fig:coupling}
\end{figure}

It is clear that the enhancement of the magnetoelectric coupling originates from the interface
where the relative energetics of spin, FE and AFD are changed compared with bulk CaMnO$_3$. The AFD 
instability is largely suppressed thus the spin-polarization coupling strength is increased. 
To further understand how the interface tunes the competetion between the AFD
and FE modes, we resort to an effective Hamiltonian. For bulk CaMnO$_3$ in a uniform phase, 
the effective Hamiltonian is, 
\begin{eqnarray}
\label{eq:Heff}
E(\{\textbf{u}_{\Gamma},\textbf{u}_{\varphi} \}) &=& E_0 -\sum_{ij}J_{ij}(0)
{\bf S}_i\cdot {\bf S}_j \\ \nonumber
&+& \alpha{\bf u}_{\Gamma}^{2}
+ \beta{\bf u}_{\varphi}^{2}+ \eta {\bf u}_{\Gamma}^{4}+ \kappa {\bf
  u}_{\varphi}^{4}+ \gamma {\bf u}_{\Gamma}^{2}{\bf u}_{\varphi}^{2}\, ,
\end{eqnarray}
where, ${\bf u}_{\Gamma}$ and ${\bf u}_{\varphi}$ are the FE and AFD modes
respectively. The phonon frequencies 
$\alpha = {1 \over 2}m_{\Gamma}
\omega^{2}_{\Gamma}-\sum_{ij}{\partial^{2}J_{ij} 
\over \partial^{2} u_{\Gamma}}{\bf S}_i\cdot {\bf S}_j$ and $\beta= {1 \over 2}m_{\varphi}
\omega^{2}_{\varphi}-\sum_{ij}{\partial^{2}J_{ij} 
\over \partial^{2} u_{\varphi}}{\bf S}_i\cdot {\bf S}_j$ strongly
depend on the spin configurations of the system.
The anharmonic terms $\kappa$, $\eta$ are positive. $\gamma$ which
describes the coupling
between the FE and AFD modes, is also positive. 
At the energy minima, we have 
${\partial E(\{\textbf{u}_{\Gamma},\textbf{u}_{\varphi})/ \partial{\bf
    u}_{\Gamma}}$=0, and 
${\partial E(\{\textbf{u}_{\Gamma},\textbf{u}_{\varphi})/ \partial{\bf
    u}_{\varphi}}$=0, i.e.,
\begin{eqnarray}
\alpha{\bf u}_{\Gamma}
+ 2 \eta {\bf u}_{\Gamma}^{3}+  \gamma {\bf u}_{\Gamma}{\bf u}_{\varphi}^{2}
=0 \\
\beta{\bf u}_{\varphi}
+ 2 \kappa {\bf u}_{\varphi}^{3}+  \gamma {\bf u}_{\varphi}{\bf u}_{\Gamma}^{2} =0
\end{eqnarray}
It is easy to see from the equations that if $\alpha >$ 0 ($\beta >$0), we
have ${\bf u}_{\Gamma}$=0 (${\bf u}_{\varphi}$=0). The interesting cases are that both  $\alpha$ and
$\beta$ are negative. When the coupling between the two modes is weak, 
i.e., $\gamma < \gamma_c = 
{\rm min} (2\kappa\alpha/\beta, 2\eta\beta/\alpha)$, one has a solution that both 
${\bf u}_{\Gamma}$ and ${\bf u}_{\varphi}$ are non-zero as schmeatically shown in Fig.~\ref{fig:coupling} (a).
However, when  $\gamma > \gamma_c$, one has ${\bf u}_{\Gamma}$=0 and ${\bf
  u}^2_{\varphi} = -\beta / 2\kappa$, if $\beta / 2\kappa < \alpha / 2\eta $, (${\bf u}_{\varphi}$=0 and ${\bf
  u}^2_{\Gamma} = -\alpha / 2\eta$, if  $\beta / 2\kappa > \alpha / 2\eta $),
i.e., one of the soft mode is fully suppressed by the other mode that has
stronger instability, because of the coupling between the two modes,
as shown in Fig.~\ref{fig:coupling} (b).
This is exactly the case in CaMnO$_3$. 
In bulk materials, the AFD mode is more unstable, 
therefore the FE mode is fully suppressed. 
In the 1:1 SL, when the MnO$_2$ intra-layer spin configuration is
AFM, the AFD instability $\alpha$ is 
still larger than that of the FE mode despite of the interface effects, 
therefore, there are still significant AFD rotations, 
even though they are smaller than those in bulk CMO. 
However, when the MnO$_2$ intra-layer spin configuration changes to FM, 
the FE instability is enhanced \cite{lee10}, 
and becomes stronger than the AFD instability 
with the help of the BaO layer pinning effects to the AFD mode.
Therefore, the AFD mode is fully suppressed. 
The suppression of the AFD mode greatly enhances the FE modes, 
again because of the
coupling between the FE and AFD modes. 
The results for the 2:2 SL can also be understood in the same scenario.

The magnetoelectric coupling effects in the CMO/BTO SLs can be
observed by various experimental techniques. For example, one should
be able to observe large polarization change by applying magnetic
field. This can be best seen near the paramagnetic (PM) to AFM phase transition
temperatures, where relative small magnetic field is needed. Or one
may simply observe the polarization change at PM to AFM transitions.
Beside the electric polarization, the dielectric constants are also
expected to change dramatically under magnetic field near the magnetic phase
transitions.

To summarize, we have demonstrated a novel mechanism that
could lead to giant magnetoferroelectric coupling in the
multiferroic CMnO$_3$/BaTiO$_3$ superlattices. 
The key idea is that the energetics of the 
instabilities, such as antiferro-distortive mode, ferroelectric mode,
and the magnetic ordering can be drastically 
modified by interface engineering to enhance the magnetoelectric coupling.
The enhancement of the magnetoelectric coupling
is due to the interface atomistic effects 
which could be much stronger than those with mechanical coupling. 
It therefore opens a new path 
to design novel multiferroic superlattices with 
strong magnetoelectric coupling.


We acknowledge X. X. Xi for useful
discussions. LH acknowledges the support from the Chinese National
Fundamental Research Program 2011CB921200 and
National Natural Science Funds for Distinguished Young Scholars. 
XW acknowledges the support by the National Science Foundation through 
TeraGrid resources provided by NICS under grant number [TG-DMR100121].


\end{document}